\documentclass{svmult}

\usepackage{graphicx}
\usepackage{makeidx}
\usepackage{multicol}
\usepackage[bottom]{footmisc}

\begin{document}
\title*{Cluster hot flow anomaly observations during solar cycle minimum} 
\author{G.~Facsk\'o\inst{1}\thanks{Now at LPCE, 3A, Avenue de la Recherche Scientifique, 45071 ORL\'EANS CEDEX 2, France} \and M.~T\'atrallyay\inst{1} \and G.~Erd\H{o}s\inst{1} \and I.~Dandouras\inst{2}}
\institute{KFKI Research Institute for Particle and Nuclear Physics, H-1525 Budapest, P.O.Box 49, Hungary, \texttt{gfacsko@cnrs-orleans.fr} \and CESR, 9, Avenue du Colonel ROCHE, 31028 TOULOUSE CEDEX 4, France}
\titlerunning{Cluster hot flow anomaly observations during solar cycle minimum}
\authorrunning{Facsk\'o et al.}
\maketitle

\begin{abstract}
Hot flow anomalies (HFAs) are studied using observations of the FGM magnetometer and the CIS plasma detector aboard the four Cluster spacecraft. Previously we studied several specific features of tangential discontinuities on the basis of Cluster measurements in February-April 2003 and discovered a new condition for forming HFAs that is the solar wind speed is higher than the average. However during the whole spring season of 2003, the solar wind speed was higher than average. In this study we analyse HFAs detected in 2007, the year of solar cycle minimum. Our earlier result was confirmed: the higher solar wind speed is a real condition for HFA formation; furthermore this constraint is independent of Schwartz et al.'s \cite{schwartz00:_condit} condition for HFA formation. 
\end{abstract}

\section{Introduction}
\label{sec:intro}

Hot flow anomalies were discovered in the 1980s \cite{schwartz85, thomsen86:_hot}. It is commonly agreed that HFAs are generated by the interaction of a tangential discontinuity and the bow shock. They are explosive events, particle acceleration, magnetic depletion take place at their center and the magnitude of the magnetic field increases at the rim. The plasma temperature increases whereas the density drops within the affected region, but the most interesting phenomenon is the directional change of the solar wind flow. The flow turns away from the anti-sunward direction and might even be directed backward \cite{sibeck99:_compr, sibeck02:_wind, thomsen93:_obser_test_hot_flow_abnom, thomsen86:_hot}. Both the previous global investigation \cite{schwartz00:_condit} and our earlier studies also indicated that HFAs are not rare phenomena \cite{kecskemety06:_distr_rapid_clust}. Cluster multispacecraft measurements help not only to understand the microphysics of HFA formation \cite{lucek04:_clust} but also to detect much more events than before because the spacecraft stay close to the bow shock for a long time; furthermore Cluster satellites have all necessary instruments to detect HFAs. 

We checked observationally a previous HFA simulation \cite{lin02:_global} determining the different angles using Cluster FGM and CIS HIA together with ACE MAG and SWEPAM measurements in the solar wind. In doing so, we discovered a new condition of HFA formation which was not predicted by the simulations; namely that the solar wind velocity is \emph{200km/s higher than average when HFA develops} \cite{facsko08:_clust}. Although a previous investigation in the magnetosheath \cite{safrankova00:_magnet} supports our result, however we want to check this condition in another year closer to the solar cycle minimum because during the whole spring season of 2003, the solar wind speed was higher than average. We chose 2007 because this is close to the solar minimum and the average solar wind velocity was expected to be lower than in 2003 when solar wind velocity was higher than the average value in the whole February-April season. We searched and analysed Cluster measurements in 2007 using the same methods like in \cite{facsko08:_clust} and compared them with previous results. 

The structure of this paper is the following: in Section~\ref{sec:obs} we discuss the processed and analysed measurements and in Section~\ref{sec:disc} we give the summary of our results. 

\section{Observations}
\label{sec:obs}

We set the same criteria for the selection of HFA events as in our previous papers \cite{facsko08:_clust, kecskemety06:_distr_rapid_clust}. Using these criteria we found more then 50 HFA events in January-April 2007 and in all cases we could determine the discontinuity normal from Cluster FGM and ACE MAG measurements. 

\begin{figure}[t]
\centering
\includegraphics[width=250pt,height=250pt]{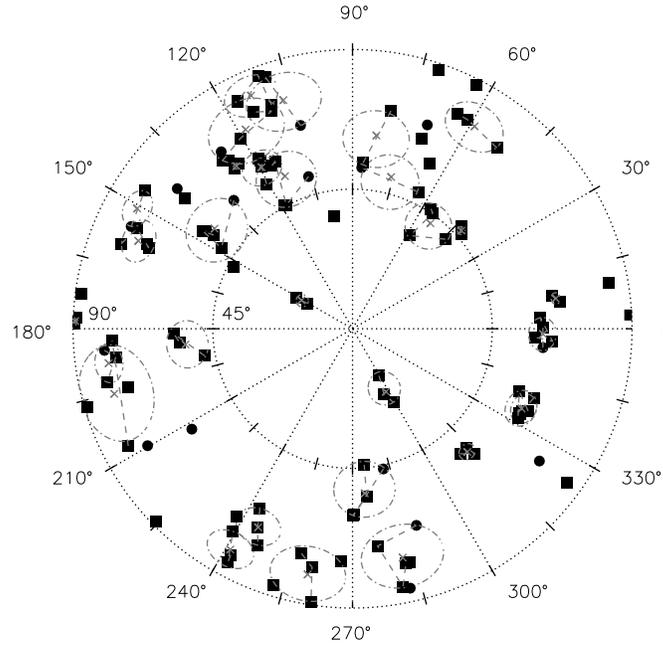}  
\caption{Polar plot of the direction of the normal vectors of TDs. The azimuthal angle is measured between the GSE y direction and the projection of the normal vector onto the GSE yz plane. The distance from the center is the cone angle as determined by the cross-product method. The regions surrounded by dashed dotted lines are the projection of error cones around the average normal vector marked by ``X''. Circles and squares symbolize ACE and Cluster data, respectively.}
\label{fig:fig1}
\end{figure}

We identified the tangential discontinuities which generated HFA events between January 1 and April 30, 2007 using Cluster-1 and -3 FGM 1s average magnetic field and CIS 4s average solar wind speed measurements as well as MAG magnetic (16s) and SWEPAM (64s) plasma data from ACE. We applied both the minimum variance techniques \cite{sonnerup98:_minim_maxim_varian_analy} and the cross product method \cite{paschmann98:_analy_method_for_multi_spacec_data} to determine their normal vectors. We checked whether the normal component of the magnetic field disappears at the discontinuity. If its value approaches zero then the discontinuity was identified as tangential and we could use the cross product method to determine its normal vector. We accepted the result of minimum variance method if the angle between the minimum variance and the crossproduct vector was less then $15^o$. We plotted all discontinuity normal determined using Cluster FGM and ACE MAG measurements. If a HFA was observed and its TD normal was calculated from more than one Cluster and the ACE spacecraft, the average vector was also plotted and the largest deviation from its direction was considered as the error and it was drawn by dashed-dotted line on Fig.~\ref{fig:fig1}. Normal belonging to the same event was connected by dashed line. 

Almost all the cone angles of the TD normals are larger than $45^o$ as predicted theoretically \cite{lin02:_global} and confirmed observationally \cite{facsko08:_clust, schwartz00:_condit}. We could not find TD-s neither in ACE nor in Cluster in the second part of March and in April (See: Fig.~\ref{fig:fig3}). We found many embedded HFAs in SLAMS in this interval and we could not determine the TD of those events. 

\begin{figure}[t]
\centering
\includegraphics[width=200pt,height=200pt]{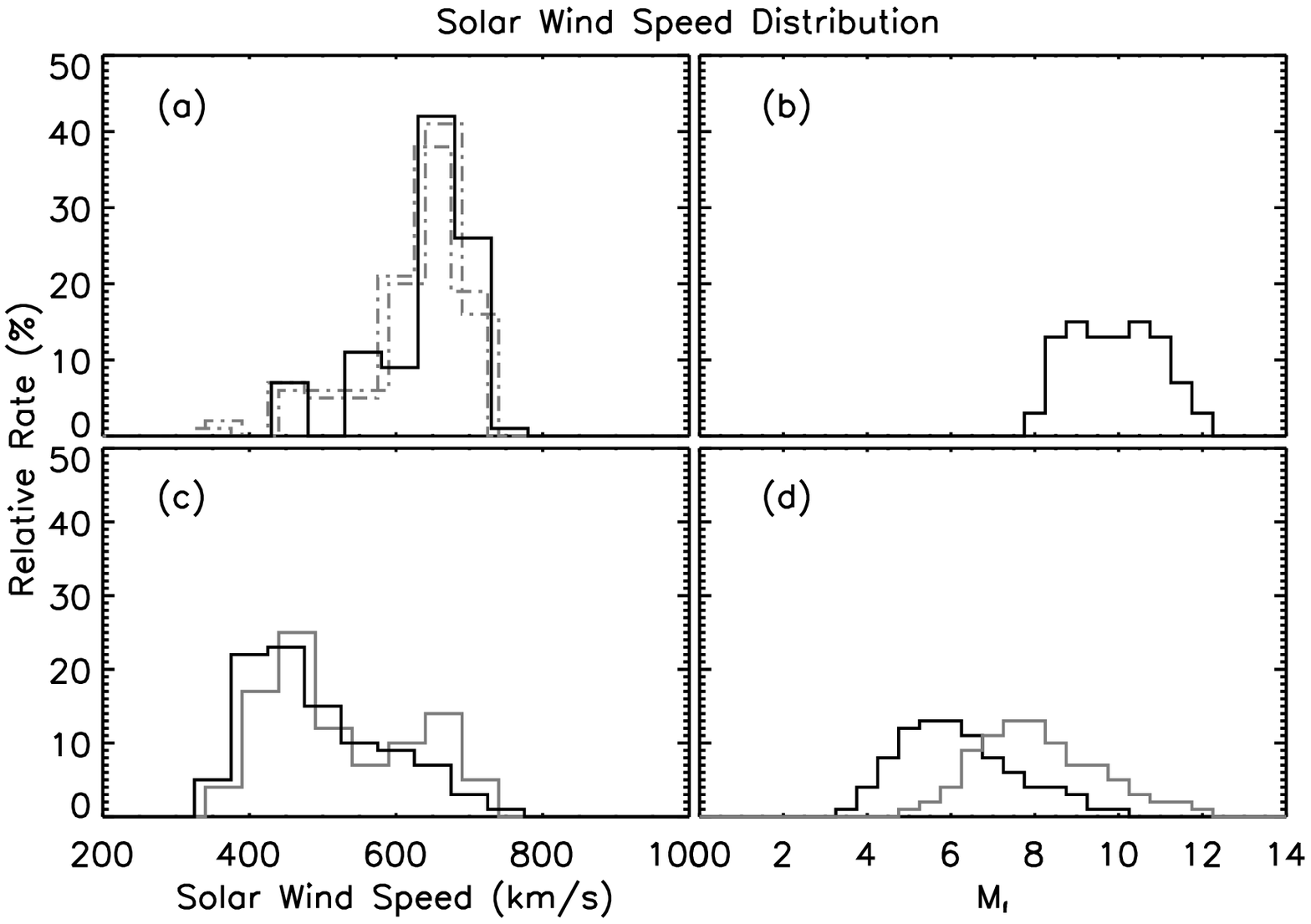}  
\caption{Upstream solar wind speed distribution measured by Cluster and ACE spacecraft. (a) Solar wind speed distribution measured by Cluster-1 (shifted grey dash dotted line) and Cluster-3 CIS HIA (shifted grey dash dotted line) ACE SWEPAM (solid line) upstream of HFA formation. (b) Fast magnetosonic Mach number distribution calculated using ACE MAG and SWEPAM data during HFA formation. (c) Solar wind speed distribution measured by ACE SWEPAM from January to April, 2007 (grey line) and from 1998 to 2008 (solid line). (d) Same as (c) but measured in $M_f$ units.}
\label{fig:fig2}
\end{figure}

\begin{table}[b]
\centering
\begin{tabular}[c]{r||rcl|rcl|c}
solar wind speed $\left(km/s\right)$ & \multicolumn{3}{c|}{2003} & \multicolumn{3}{c|}{2007} & Fig \\
\hline
during HFA formation by C1 & $680$&$\pm$&$86$ & $613$&$\pm$&$80$ & \ref{fig:fig2}a \\
 by C3 & $671$&$\pm$&$92$ & $613$&$\pm$&$78$ & \\
 by ACE & $666$&$\pm$&$84$ & $634$&$\pm$&$71$ & \\
$M_f$ numbers by ACE & $8.2$&$\pm$&$1.2$ & $9.9$&$\pm$&$1.1$ & \ref{fig:fig2}b \\
\hline
in 3/4 months period by ACE & $546$&$\pm$&$97$ & $512$&$\pm$&$102$ & \ref{fig:fig2}c \\
\hline
between 1998-2003/2008 by ACE & $492$&$\pm$&$102$ & $498$&$\pm$&$101$ & \ref{fig:fig2}c \\
$M_f$ numbers by ACE & $5.5$ & $\pm$ & $1.4$ & $6.2$ & $\pm$ & $1.7$ & \ref{fig:fig2}d \\
\hline \hline
$\Delta M_f$ & \multicolumn{3}{c|}{$2.7$} & \multicolumn{3}{c|}{$3.7$} & \\
\end{tabular}
\caption{Solar wind speed and fast magnetosonic Mach numbers mean values and their deviation measured by Cluster CIS and ACE SWEPAM. The last column gives the figure numbers on Fig.~\ref{fig:fig2}.}
\label{tab:sw}
\end{table}

We determined the solar wind velocity and fast-magnetosonic Mach-number distributions upstream of HFA formation and compared them to the same distributions in spring 2007 and for about 10 years (Fig.~\ref{fig:fig2}, Tab.~\ref{tab:sw}). The average velocities seem to be lower than in 2003 but with error it is not so significant. Although Mach number average values seem to be higher in 2007 but its error is too high and makes this difference not so significant. This means that the solar wind velocity is higher when HFA develops but the difference is only 120\,km/s in 2007. The difference measured in Mach numbers is greater \cite{facsko08:_clust}, but the difference is within the error limit. In Fig.~\ref{fig:fig2}c the typical double peak distribution of solar wind velocity \cite{erd05:_in} can be seen for spring 2007. It helps us to understand why the solar wind distribution is anomalous during HFA formation on Fig.~\ref{fig:fig2}a. After comparing the positions of the maximum on Fig.~\ref{fig:fig2}b and Fig.~\ref{fig:fig2}d the difference in Mach numbers can be seen.  

\begin{figure}[t]
\centering
\includegraphics[width=200pt]{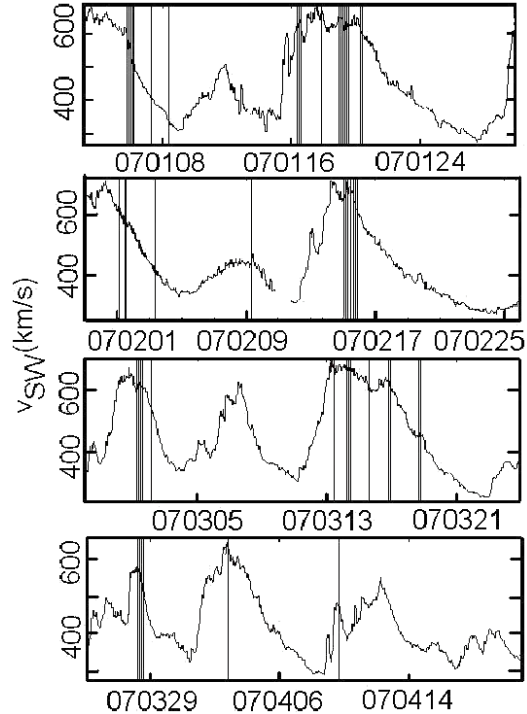}  
\caption{The measured SW speed at the L1 point for 4 solar rotations. ACE 1-hr averages are from ACE Science Center. The times of the observed HFAs are marked by vertical lines.}
\label{fig:fig3}
\end{figure}

We also plotted the $1^h$ averaged solar wind speed time series from January to April, 2007 (Fig.~\ref{fig:fig3}). The last HFA event, on April 30, 2007 was not plotted, but it is evident that HFA events are coupled to high solar wind velocity intervals like in spring 2003. 

\section{Discussion and conclusions}
\label{sec:disc}

Our final conclusions based on measurements in spring 2007 are the following:
\begin{enumerate}
\item The solar wind velocity is about 120\,km/s or $\Delta M_f=3.7$ higher than the average when HFA develops.  
\item HFA events occur almost always when the solar wind speed is higher than average. 
\item The angle between the solar direction and the identified TD normal is usually greater then $45^o$ like in 2003.
\item Approximately another 40-50 HFA events were embedded in SLAMS, but the TD raising them could not be identified.
\end{enumerate}
Based on observations it is hard to say whether the higher solar wind velocity or Mach number is more important. Further independent theoretical considerations \cite{nemeth07:_partic} suggest that both higher Mach number and higher solar wind speed are important in HFA formation. This theoretical work studied the orbit of particles accelerated by the bow-shock and led back to the shock by the convective electric field ($-\mathbf{v}\times\mathbf{B}$) on both side of the tangential discontinuity. The particles are accelerated -- and form a HFA - if the angle between the solar wind direction and the TD normal is less than the calculated critical angle. This critical angle depends on the ratio of the upstream and downstream magnetic field (which is a function of Mach number) and the solar wind speed. Furthermore this critical angle is $41.8^o$ calculated for typical Mach numbers and high solar wind speed so the theory confirms our observational results and the prediction of Lin's computational work \cite{lin02:_global}. 

\begin{figure*}[t]
\centering
\includegraphics[width=150pt]{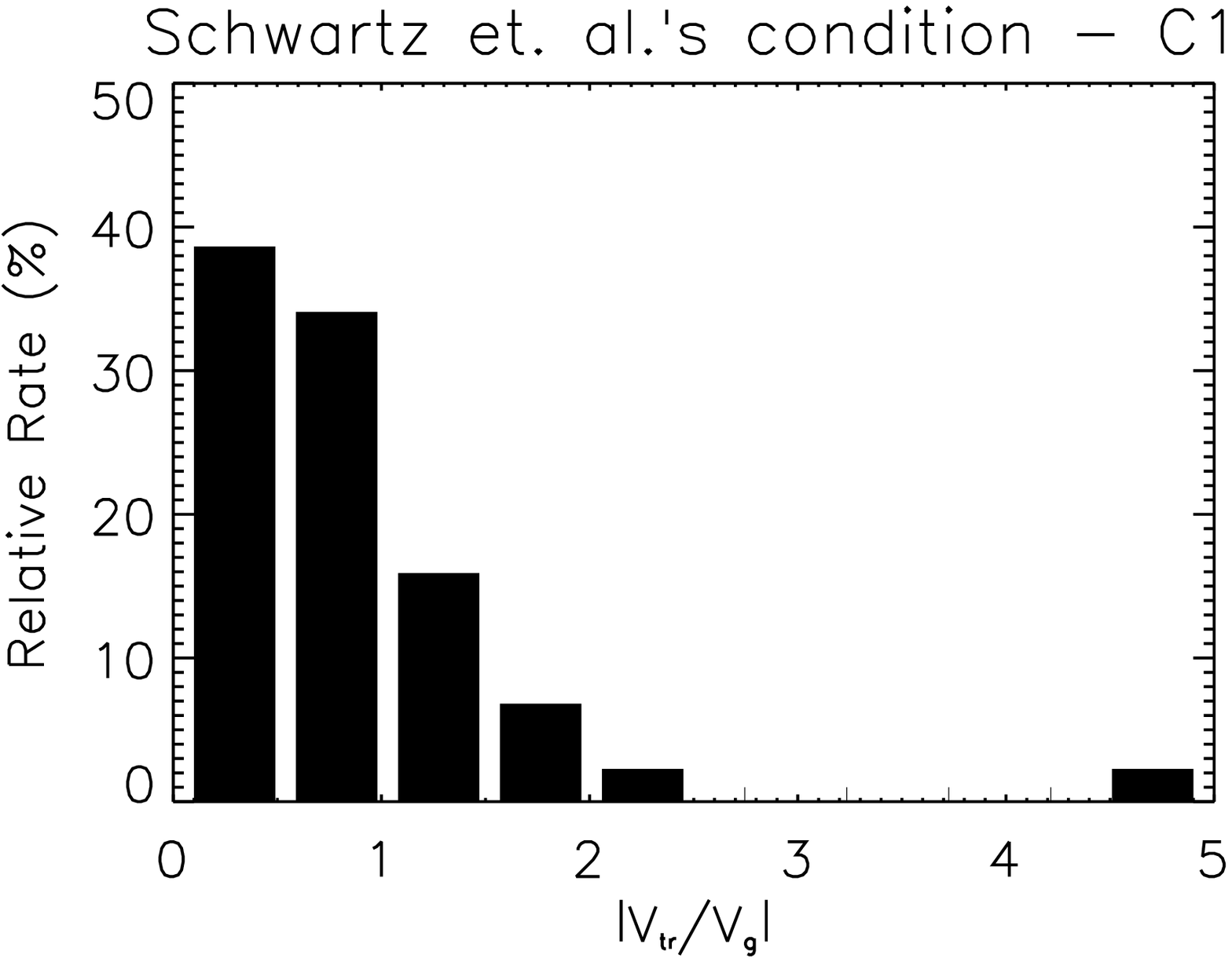}  
\includegraphics[width=150pt]{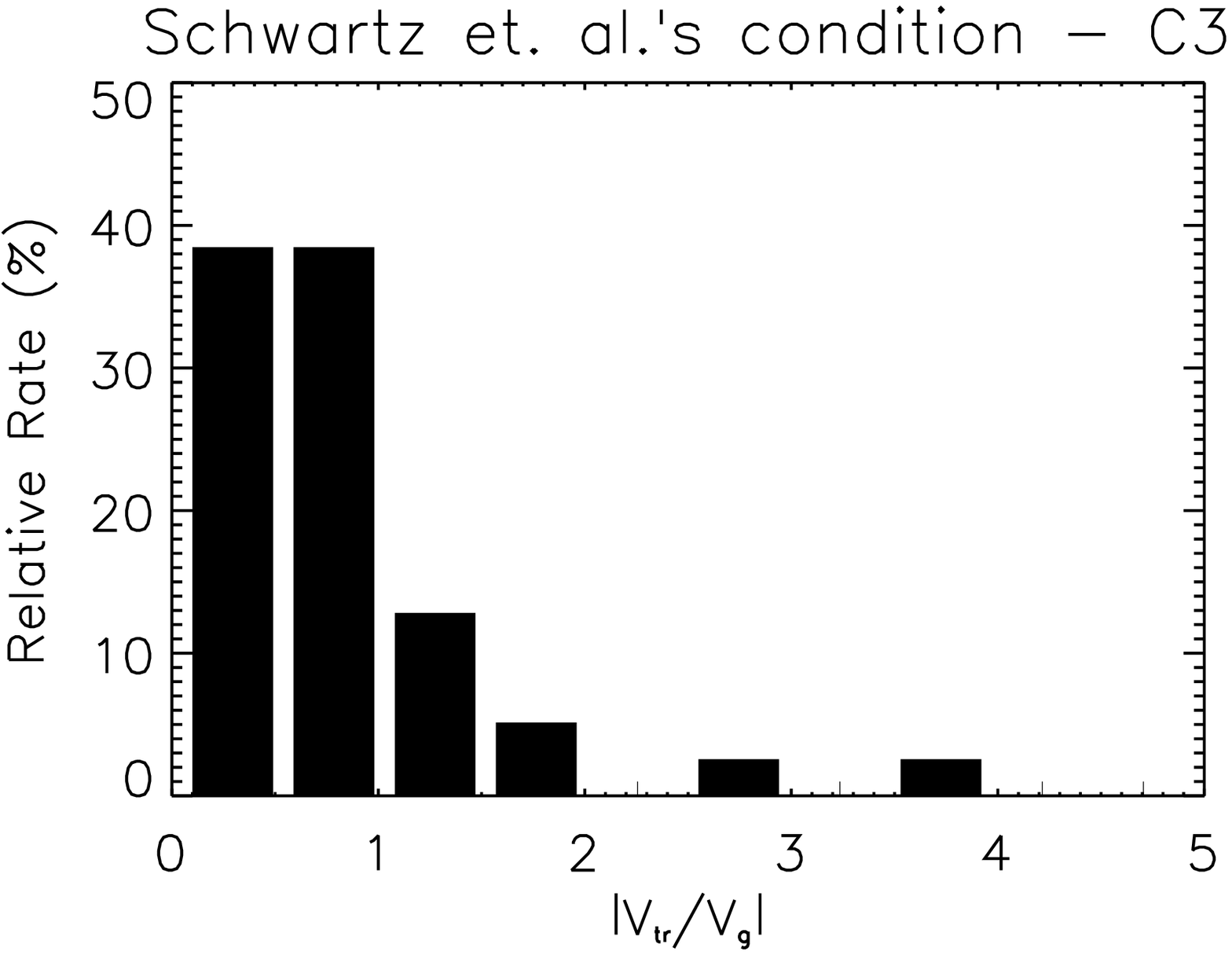}    
\caption{The distribution of the value of Schwartz el al.'s formula \cite{schwartz00:_condit} by Cluster-1 and -3 measurements. The values are mostly less than 1.0. }
\label{fig:schwartz}
\end{figure*}

The above discussed analysis of HFA events in the spring 2007 season confirmed and extended our earlier results based on the study of HFA events in spring 2003 that higher solar wind speed is an important condition of HFA formation. This feature restricts Schwartz et al.'s formula \cite{schwartz00:_condit} but these events also confirm their result:

\begin{eqnarray*}
\label{eq:schwartz}
\left|\frac{V_{tr}}{V_g}\right| &=& \frac{\cos\theta_{cs:sw}}{2\cos\theta_{bs:sw}\sin\theta_{B_n}\sin\theta_{cs:bs}}, \\
\end{eqnarray*}
where $V_{tr}$ is the transit velocity of the current sheet along the bow-shock, $V_g$ is the gyration speed, $\theta_{cs:sw}$, $\theta_{bs:sw}$ and $\theta_{cs:bs}$ are the angles between the discontinuity normal, solar wind velocity and the bow-shock; furthermore $\theta_{B_n}$ is the angle between the magnetic field and bow-shock normal. The necessary vectors were calculated and we got that the transition speed is as low as expected by Schwartz et al.'s formula \cite{schwartz00:_condit} (Fig.~\ref{fig:schwartz}). So, in addition to demonstrating that HFA formation depends on solar wind speed, this study confirms previous work predicting that HFA formation also depends on the geometry of the shock relative to the solar wind. 

\section*{Acknowledgements}
\label{sec:acknow}

The authors thank the OTKA grant K75640 of the Hungarian Scientific Research Fund for support and the Cluster FGM, CIS and ACE MAG, SWEPAM teams for providing data for this study. 

\bibliography{gfacsko_paper}
\bibliographystyle{plain}

\end{document}